\documentclass[document,showpacs]{revtex4}
\pdfoutput=1
\usepackage{amssymb,latexsym}
\usepackage{amsmath,amsbsy,bbm}
\usepackage{ifpdf}
\usepackage{epsfig,bm}
\usepackage{graphicx,comment}
\usepackage{color}
\usepackage{soul}
\usepackage{mathtools}
\usepackage{comment}
\usepackage[normalem]{ulem}
\begin{document}

\title{Neural network evidence of a weakly first order phase transition for
the two-dimensional 5-state Potts model}
\author{Yuan-Heng Tseng}
\affiliation{Department of Physics, National Taiwan Normal University,
  88, Sec.4, Ting-Chou Rd., Taipei 116, Taiwan}
\author{Yun-Hsuan Tseng}
\affiliation{Department of Physics, National Taiwan Normal University,
  88, Sec.4, Ting-Chou Rd., Taipei 116, Taiwan}
\author{Fu-Jiun Jiang}
\email[]{fjjiang@ntnu.edu.tw}
\affiliation{Department of Physics, National Taiwan Normal University,
88, Sec.4, Ting-Chou Rd., Taipei 116, Taiwan}

\begin{abstract}
	A universal (supervised) neural network (NN), which is only trained once on a one-dimensional lattice of 200 sites, is employed to
        study the phase transition of the two-dimensional (2D) 5-state ferromagnetic Potts model on the square lattice. In particular, the NN
        is obtained by using merely two
        artificially made configurations as the training set. Due to the elegant features of the
        considered NN, results associated with systems consisting of over
        4000000 spins can be obtained with ease, and convincing evidence showing the investigated phase transition is weakly
        first order is reached. The outcomes demonstrated here can hardly be
        achieved with the
        standard NNs that are commonly used in the literature.

\end{abstract}


\maketitle

\section{Introduction}
\label{sec:intro}

Phase transitions typically can be
classified with two categories, namely the first-order and the second-order
phase transitions (Here we will not discuss more sophisticated division
of phase transitions). Specifically, if the order parameter continuously
goes to zero when one is approaching the critical point (CP), then the related
phase transition is of second order. On the other hand, if
the value of the order parameter drops to zero suddenly as one reaches the CP,
then the corresponding phase transition is of first order.
The correlation length diverges at CP for a system which undergoes a second-order phase transition,
while it remains finite for a first-order phase transition.
The identification of a weakly first-order phase transition is a challenge.
This is because it has a very long correlation length, and the system size
must be much larger than the correlation length in order to capture the true
physics of the considered phase transition.

Recently, the machine learning (ML) techniques are applied to
study many-body systems \cite{Li15,Mnih:2015jgp,Baldi:2016fzo,Searcy:2015apa,Baldi:2016fql,Wan16,Car16,Tro16,Chn16,Bro16,Nie16,Tubiana:2016zpw,Kol17,Liu16,Wan17,Liu17,Nag17,Xu16,Hu17,Mott:2017xdb,Pang:2016vdc,Li18,Bar18,Zha18,Gao18,Lu18,But18,Lia19,Rod19,Conangla:2018nnn,Zha19,Gre19,Car19,Meh19,Don19,Tan20.1,Bachtis:2020dmf,Beentjes:2020abj,Tan20.2,Sch20,Lidiak:2020vgk,Carrasquilla:2020mas,Baldi:2014pta,Han:2019wue,Larkoski:2017jix,Shalloo:2020nhu,Aad:2020cws,Nicoli:2020njz}. In particular, the neural network (NN) methods
have been demonstrated to be able to identify various phases of
many physical models including the Ising model, the Potts models,
the $O(N)$ models, and the Hubbard type models \cite{Car16,Chn16,Bro16,Li18,Tan20.1,Tan20.2}.

While satisfactory outcomes are obtained, some bottlenecks of employing the NN approach
to investigate many-body systems remain. One noticeable example is to determine the nature of weakly
first order phase transitions definitely. Indeed, the standard NN calculations,
particularly in the training stages, require huge amount of computing resources. As a result,
for moderate computing power, most of the associated NN calculations are limited to small to intermediate
system sizes.

To overcome the issues typically encountered in NN calculations, a one-dimensional (1D) NN was constructed in Ref.~\cite{Tan21}.
It is shown that the needed computing power for carrying the related investigation using the built 1D NN of Ref.~\cite{Tan21}
is much less than that of the standard NN approach. In particular, when compared with the standard NN schemes,
several hundred to few thousand factors in efficiency and storage is gained for the 1D NN.
Moreover, the mentioned 1D NN is universal because it has successfully
determined the CPs of many three-dimensional (3D) and two-dimensional (2D) models. 

In order to further demonstrate the power and the advantage of the 1D NN of Ref.~\cite{Tan21}, in this study we employed that 1D
NN to investigate the phase transition of the two-dimensional (2D) 5-state ferromagnetic Potts model
on the square lattice. Due to the elegant features of the constructed 1D NN, here we are able to
study systems with more than 4000000 spins. As a result, convincing evidence showing that the studied phase transition is
(weakly) first order is obtained.

With moderate computing resources, the calculations carried here can hardly be achieved by
any of the standard NN approaches available in the literature.

\section{The constructed universal supervised Neural Network}

\begin{figure*}
        \begin{center}
                \includegraphics[width=0.95\textwidth]{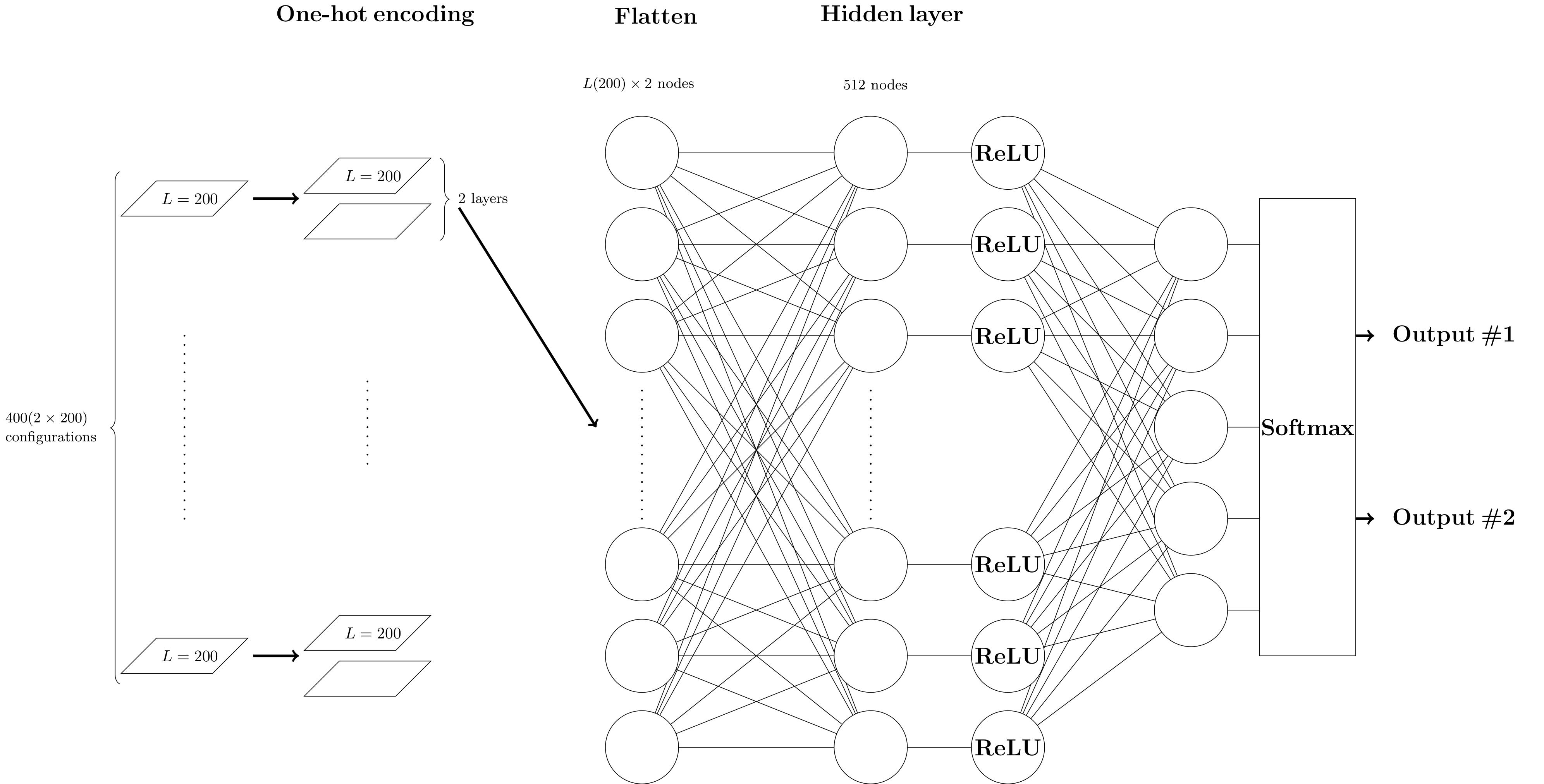}
        \end{center}\vskip-0.7cm
        \caption{The NN (MLP) used here and in Refs.~\cite{Tan20.1,Tan20.2,Tan21}.
                The training set are
                made up of 200 copies of only two artifically made configurations.
                The steps of one-hot encoding and flattening are applied. ReLU and softmax
                are the employed activation functions.
                The output layer consists of two-component vector(s).
                The figure is reproduced from Refs.~\cite{Tan20.1,Tan20.2,Tan21}}
        \label{MLP}
\end{figure*}

Using the publically available libraries keras and tensorflow \cite{kera,tens},
the 1D (supervised) NN employed here is a multilayer perceptron (MLP) which consists
of only one input layer, one hidden layer of 512 
independent nodes, and one output layer. The training set are
two artificially made one-dimensional (1D) lattice of 200 sites.
In addition, all the sites of one configuration for training take 1 as their values
and each element of the other (configuration) is 0. The used labels
are $(1,0)$ and $(0,1)$. A thorough introduction for the infrastructure of the
considered NN and the training as well as the testing (prediction) procedures are
available in Refs.~\cite{Tan20.1,Tan21}, and fig.~\ref{MLP} is the associated cartoon representation
of the 1D NN employed in this study \cite{Tan21}. Here we will not go into the details of these terminologies
and processes, and only would like to emphasize the fact that no NN is trained
in this investigation. In other words, the employed NN in this study is adopted directly from Ref.~\cite{Tan21}.
For the readers who are interested in the fundamental details of applying NN to investigate physical systems
are referred to Refs.~\cite{Car19,Meh19}

\section{The microscopic model and the observables}

The Hamiltonian of the studied 2D 5-state ferromagnetic Potts model on the square lattice
is given by \cite{Wu82,Swe87} 
\begin{equation}
\beta H_{\text{Potts}} = -\beta \sum_{\left< ij\right>} \delta_{\sigma_i,\sigma_j},
\label{eqn1}
\end{equation}
where $\beta$ is the inverse temperature and $\left< ij \right>$ stands for
the nearest neighbor sites $i$ and $j$. In addition, in Eq.~(\ref{eqn1})
$\delta$ refers to the Kronecker function and the Potts variable $\sigma_i$
at each site $i$ takes an integer value from $\{1,2,3,4,5\}$.
The observable considered here
is the energy $E$.

As the temperature ($T$) changes from low temperatures to high temperatures,
a phase transition will take place for the studied model.
Moreover, it is well established theoretically that the  
related critical point $T_c$ is given by  $T_c \sim 0.85153$ and the nature
of this phase transition is weakly first order \cite{Wu82}.

\section{Numerical Results}

\begin{figure}
        \begin{center}

                        \includegraphics[width=0.45\textwidth]{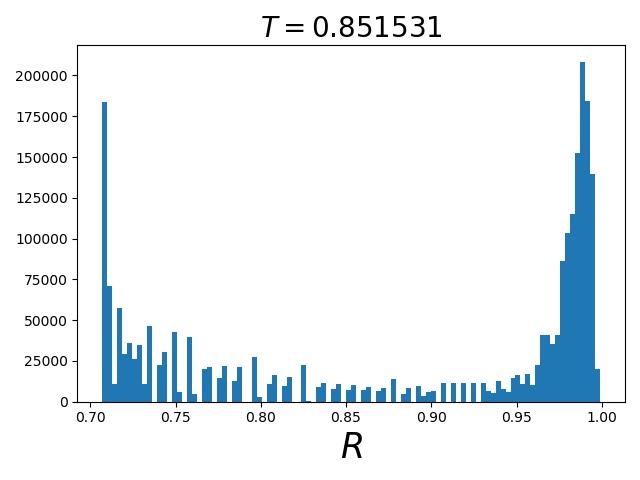}
        \end{center}\vskip-0.7cm
        \caption{The histogram of the magnitude $R$ of the NN output vectors for the 2D 5-state Potts model.
          The associated temperature $T$ and the linear box size $L$
          are given by $T = 0.851531$ and $L = 2048$, respectively. A two peak structure appears in the histogram
          clearly.}
        \label{his_5}
\end{figure}

In this section the outcomes of applying the 1D NN to study the nature of the phase transition
of the considered model are presented.
In particular, we will provide convincing NN evidence to show that the studied phase transition
is indeed (weakly) first order. The associated Monte Carlo simulations are done using the Wolff
algorithm \cite{Wol89}. In addition, for every generated configuration (a configurations is generated after
every 10 Monte Carlo sweeps), the spins of
2000 randomly chosen sites are stored, and these spins will be used to build the needed configurations for
the NN testing. Finally,
each of the histograms shown below consists of few million data points and is
produced using the histogram of pylab with 100 bins.

Theoretically, it is predicted that the phase transition of the 2D 5-state ferromagnetic Potts model
on the square lattice is weakly first order. In other words, the associated correlation length
is very large. As a result, data points on large lattices as well as
long simulations are needed in order to observe any signal
of the predicted first-order phase transition.

The histogram of the magnitude $R$ of the NN output vectors is shown in fig.~\ref{his_5}. The outcomes
in the figure are obtained with $L=2048$ and $T = 0.851531$. Clearly a two peak
structure appears
in fig.~\ref{his_5}. This provides strong evidence that the associated phase transition is of
first order. 

\begin{figure}
  \begin{center}
    \vbox{
\hbox{~~~~~~~~~~
  \includegraphics[width=0.4\textwidth]{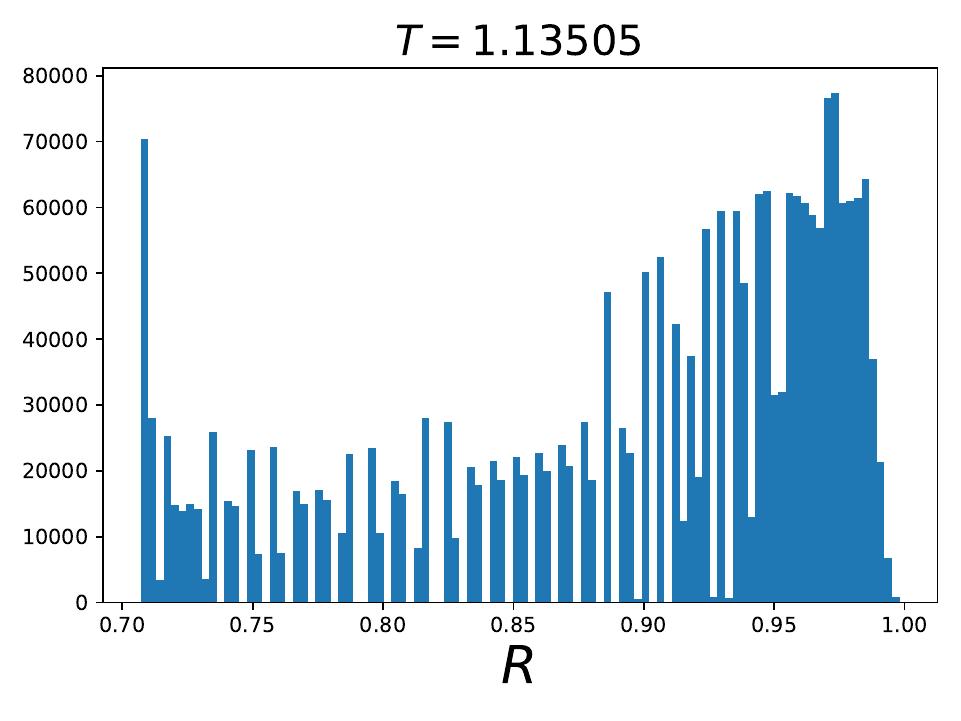}~~~~~~~~~
  \includegraphics[width=0.4\textwidth]{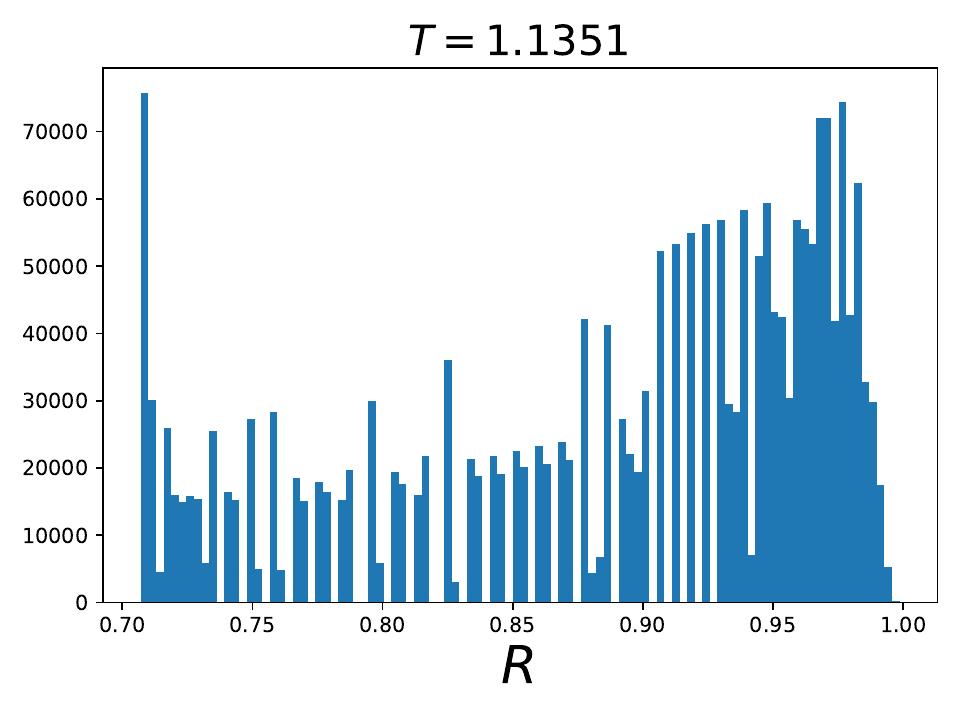}
}
\hbox{~~~~~~~~~~
  \includegraphics[width=0.4\textwidth]{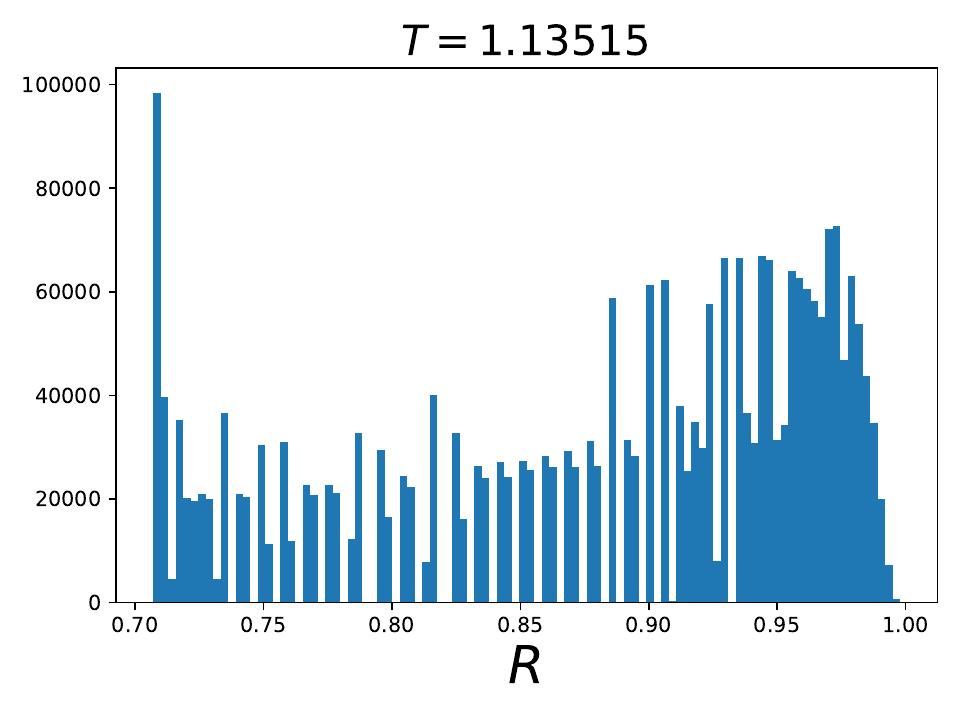}~~~~~~~~~
  \includegraphics[width=0.4\textwidth]{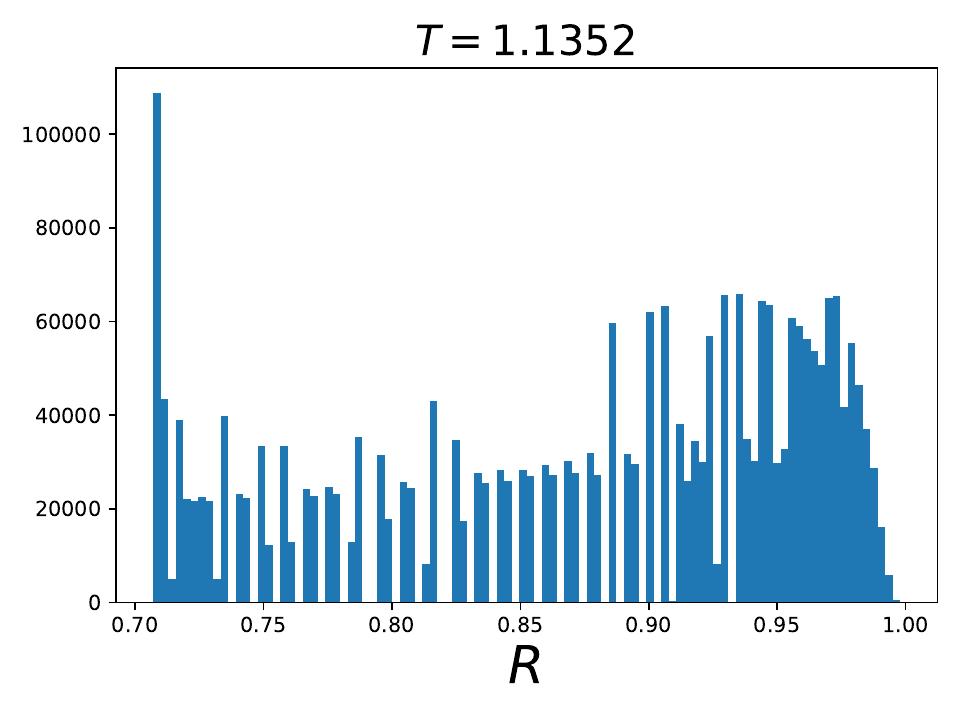}
}
}
        \end{center}\vskip-0.7cm
        \caption{The histograms of the magnitude $R$ of the NN output vectors for the 2D two-state Potts model.
          The box sizes for all panels are $L=2048$.}
        \label{his_2}
\end{figure}

To compare the difference between the histograms of $R$ for a weakly first-order and a second-order
phase transitions, we have also simulated the 2D two-state ferromagnetic Potts model on the square lattice.
It is known that for this model (two-state Potts model) the corresponding critical temperature $T_c$
is given by $T_c \sim 1.134593$. Therefore, the related calculations of $L=2048$
are carried out with $T$ = 1.1348, 1.1349, 1.1350, 1.3505, 1.1351, 1.13515, 1.1352 ,1.13525, 1.1353, 1.1354, 1.1355. Among
the NN results of these simulated $T$ values, no clear two peak structure is found, see
fig.~\ref{his_2} for some of the associated results. This outcome
confirms that the two peak structure found in fig.~\ref{his_5} is indeed an evidence of (weakly)
first-order phase transition for the 2D 5-state Potts model.

\begin{figure}
        \begin{center}

                        \includegraphics[width=0.45\textwidth]{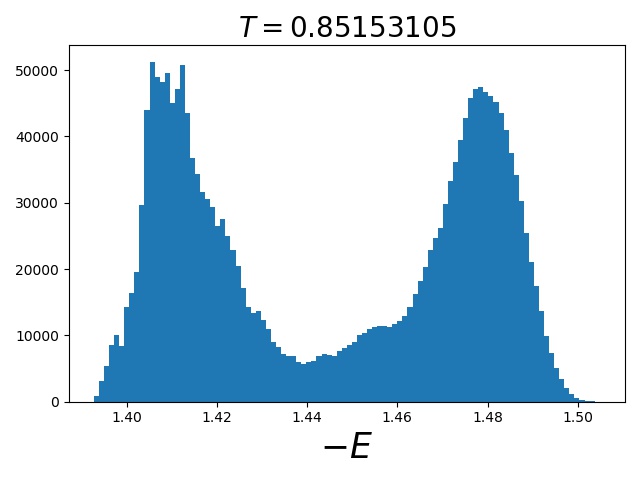}
        \end{center}\vskip-0.7cm
        \caption{The histogram of $-E$ for the 2D 5-state Potts model.
          The associated temperature $T$ and the linear box size $L$
          are give by $T = 0.85153105$ and $L = 2048$, respectively. A two peak structure appears as well in the histogram.}
        \label{his_E}
\end{figure}

Interestingly, the histogram of the quantity (minus) energy $-E$ is shown in fig.~\ref{his_E}.
By a comparison between the outcomes demonstrated in figs.~\ref{his_5} and \ref{his_E}, one
finds that the NN results have more profound two peak structure than that of $-E$. In other words,
larger lattice (than $L=2048)$ data of $-E$ are required in order to obtain stronger evidence
to show that the phase transition is first order. This task will take much more computing time.
In addition, we also find that the tunneling phenomenon of $R$ happens much more frequently
than that of $-E$.
In summary, the outcomes shown in figs.~\ref{his_5} and \ref{his_E} as well as the observed
tunneling frequencies may lead to the conclusion
that when weakly first order phase transition is concerned, the performance of NN is better
than that of the traditional approaches.

\section{Discussions and Conclusions}

In this study, we use the 1D NN of 200 sites that was only trained once in Ref.~\cite{Tan21} to investigate the nature of the phase transition of
the 2D 5-state ferromagnetic Potts model on the square lattice. Due to the elegant features of the employed NN, we are able to access systems
of over 4000000 spins with no difficulty. Such huge system sizes cannot be easily handled using the conventional NN approaches available in
the literature with moderate computing resources.

Each of fhe figures demonstrated here uses few million data points. With the conventional NN procedures,
the storage space needed for such a huge amount of data is more than $10^{13}$ bytes, which cannot be easily
managed. Only with the elegant idea used in Ref.~\cite{Tan21} and here can one reach the outcomes presented in this study with ease. 

Our NN outcomes provide convincing evidence that the targeted phase transition is weakly first order. This is consistent with the
associated theoretical prediction. In addition, the comparison between the histograms of the relevant quantities obtained by the NN and the
traditional methods implies that the NN may have better performance.

It is remarkable that the simple 1D NN of Ref.~\cite{Tan21} can be applied to study the phase transitions of many higher-dimensional systems.
These tasks can hardly be achieved by any single conventional NN established in the literature.

\section*{Data Availability Statement}
Data are available upon reasonable request to the corresponding author.

\section*{Author Contributions}
F.J.J proposed and supervised the project, and wrote up the manuscript. Y.H.T. and Y.H.T. conducted the calculations and analyzed the data.

\section*{Funding}
Partial support from Ministry of Science and Technology of Taiwan is 
acknowledged.

\section*{Conflict of Interest}
The authors declare no conflict of interest.


\begin{thebibliography}{1}


\bibitem{Li15}
Zhenwei Li, James R. Kermode, and Alessandro De Vita,
Phys. Rev. Lett. {\bf 114}, 096405 (2015).



\bibitem{Mnih:2015jgp}
V.~Mnih, K.~Kavukcuoglu, D.~Silver, A.~A.~Rusu, J.~Veness, M.~G.~Bellemare, A.~Graves, M.~Riedmiller, A.~K.~Fidjeland, G.~Ostrovski, S.~Petersen, C.~Beattie, A.~Sadik, I.~Antonoglou, H.~King, D.~Kumaran, D.~Wierstra, S.~Legg and D.~Hassabis, 
Nature \textbf{518}, no.7540, 529-533 (2015).

\bibitem{Baldi:2016fzo}
P.~Baldi, K.~Cranmer, T.~Faucett, P.~Sadowski and D.~Whiteson,
Eur. Phys. J. C \textbf{76}, no.5, 235 (2016).

\bibitem{Searcy:2015apa}
J.~Searcy, L.~Huang, M.~A.~Pleier and J.~Zhu,
Phys. Rev. D \textbf{93}, no.9, 094033 (2016).


\bibitem{Baldi:2016fql}
P.~Baldi, K.~Bauer, C.~Eng, P.~Sadowski and D.~Whiteson,
Phys. Rev. D \textbf{93}, 094034 (2016).

\bibitem{Wan16}
  Lei Wang, 
  Phys. Rev. B {\bf 94}, 195105 (2016).
  

\bibitem{Car16}
  Juan~Carrasquilla, Roger~G.~Melko,
  Nature Physics {\bf 13}, 431–434 (2017).

\bibitem{Tro16}
  Giuseppe Carleo, Matthias Troyer,
  Science 355, 602 (2017).


\bibitem{Chn16}
Kelvin Ch'ng, Juan Carrasquilla, Roger G. Melko, and Ehsan Khatami,
Phys. Rev. X {\bf 7}, 031038 (2017).

\bibitem{Bro16}
Peter Broecker, Juan Carrasquilla, Roger G. Melko, and Simon Trebst,
Scientific Reports {\bf 7}, 8823 (2017).


\bibitem{Nie16}
Evert P.L. van Nieuwenburg, Ye-Hua Liu, Sebastian D. Huber,
Nature Physics {\bf 13}, 435–439 (2017).

\bibitem{Tubiana:2016zpw}
J.~Tubiana and R.~Monasson,
Phys. Rev. Lett. \textbf{118}, 138301 (2017).



\bibitem{Kol17}
Kolb, B., Lentz, L.~C. and Kolpak, A.~M.,
Sci Rep {\bf 7}, 1192 (2017).

\bibitem{Liu16}
Junwei Liu, Huitao Shen, Yang Qi, Zi Yang Meng, Liang Fu,
Phys. Rev. B {\bf 95}, 241104(R) (2017).


\bibitem{Wan17}
Li Huang and Lei Wang, Phys. Rev. B {\bf 95}, 035105 (2017).



\bibitem{Liu17}
  Junwei Liu, Yang Qi, Zi Yang Meng, Liang Fu,
  Phys. Rev. B {\bf 95}, 041101(R) (2017).


\bibitem{Nag17}
Yuki Nagai, Huitao Shen, Yang Qi, Junwei Liu, and Liang Fu
Phys. Rev. B {\bf 96} 161102 (2017).

\bibitem{Xu16}
Xiao Yan Xu, Yang Qi, Junwei Liu, Liang Fu, Zi Yang Meng,
Phys. Rev. B {\bf 96}, 041119(R) (2017).


\bibitem{Hu17}
Wenjian Hu, Rajiv R. P. Singh, and Richard T. Scalettar, 
Phys. Rev. E {\bf 95}, 062122 (2017).

\bibitem{Mott:2017xdb}
A.~Mott, J.~Job, J.~R.~Vlimant, D.~Lidar and M.~Spiropulu,
Nature \textbf{550}, no.7676, 375-379 (2017).

\bibitem{Pang:2016vdc}
L.~G.~Pang, K.~Zhou, N.~Su, H.~Petersen, H.~Stöcker and X.~N.~Wang,
Nature Commun. \textbf{9}, no.1, 210 (2018).

\bibitem{Li18}
  C.-D. Li, D.-R. Tan, and F.-J. Jiang,
  Annals of Physics, 391 (2018) 312-331.


\bibitem{Bar18}
Albert P. Barto\'k, James Kermode, Noam Bernstein, and Ga\'bor Csa\'nyi,
Phys. Rev. X {\bf 8}, 041048 (2018).

  
\bibitem{Zha18}
  Pengfei Zhang, Huitao Shen, and Hui Zhai,
Phys. Rev. Lett. {\bf 120}, 066401 (2018).

\bibitem{Gao18}
  Jun Gao et al. 
  Phys. Rev. Lett. {\bf 120}, 240501 (2018).
  
\bibitem{Lu18}
Lu, S., Zhou, Q., Ouyang, Y. et al.,
Nat Commun 9, 3405 (2018).


\bibitem{But18}
Keith T. Butler, Daniel W. Davies, Hugh Cartwright, Olexandr Isayev, and Aron Walsh, 
Nature {\bf 559}, 547–555 (2018).
  

\bibitem{Lia19}
  Wenqian Lian {\it et al.}
  Phys. Rev. Lett. {\bf 122}, 210503 (2019). 

  
\bibitem{Rod19}
  Joaquin F. Rodriguez-Nieva and Mathias S. Scheurer,
  Nat. Phys. 15, 790–795 (2019).

\bibitem{Conangla:2018nnn}
G.~P.~Conangla, F.~Ricci, M.~T.~Cuairan, A.~W.~Schell, N.~Meyer and R.~Quidant,
Phys. Rev. Lett. \textbf{122}, 223602 (2019)

\bibitem{Zha19}
Wanzhou Zhang, Jiayu Liu, and Tzu-Chieh Wei,
Phys. Rev. E {\bf 99}, 032142 (2019).


\bibitem{Gre19}
Jonas Greitemann, Ke Liu, and Lode Pollet, 
Phys. Rev. B {\bf 99}, 060404(R) (2019).


\bibitem{Car19}
  Giuseppe Carleo, Ignacio Cirac, Kyle Cranmer, Laurent Daudet, Maria Schuld, Naftali Tishby,
  Leslie Vogt-Maranto, and Lenka Zdeborov\'a,
  Rev. Mod. Phys. {\bf 91}, 045002 (2019).


  

\bibitem{Meh19}
  Pankaj Mehta, Marin Bukov, Ching-Hao Wang, Alexandre G.R. Day,
  Clint Richardson, Charles K. Fisher, and David J. Schwab,
  Phys. Rep. 810, (2019) 1-124.

  


\bibitem{Don19}
Xiao-Yu Dong, Frank Pollmann, and Xue-Feng Zhang,
Phys. Rev. B {\bf 99}, 121104(R) (2019).


\bibitem{Tan20.1}
  D.-R. Tan { \it et al.}
  2020 New J. Phys. 22 063016. 

\bibitem{Bachtis:2020dmf}
D.~Bachtis, G.~Aarts and B.~Lucini,
Phys. Rev. E \textbf{102}, no.3, 033303 (2020).

\bibitem{Beentjes:2020abj}
S.~V.~Beentjes and A.~Khamseh,
Phys. Rev. E \textbf{102}, no.5, 053314 (2020).
  
\bibitem{Tan20.2}
  D.-R. Tan and F.-J. Jiang,
  Phys. Rev. B {\bf 102}, 224434 (2020).
  
  
\bibitem{Sch20}
Mathias S. Scheurer and Robert-Jan Slager,
Phys. Rev. Lett. {\bf 124}, 226401 (2020).

\bibitem{Lidiak:2020vgk}
A.~Lidiak and Z.~Gong,
Phys. Rev. Lett. \textbf{125}, no.22, 225701 (2020).

\bibitem{Carrasquilla:2020mas}
J.~Carrasquilla,
Adv. Phys. X \textbf{5}, no.1, 1797528 (2020).


\bibitem{Baldi:2014pta}
P.~Baldi, P.~Sadowski and D.~Whiteson,
Phys. Rev. Lett. \textbf{114}, 111801 (2015).



\bibitem{Han:2019wue}
X.~Han and S.~A.~Hartnoll,
Phys. Rev. X \textbf{10}, 011069 (2020).


\bibitem{Larkoski:2017jix}
A.~J.~Larkoski, I.~Moult and B.~Nachman,
Phys. Rept. \textbf{841}, 1-63 (2020).


\bibitem{Shalloo:2020nhu}
R.~J.~Shalloo, S.~J.~D.~Dann, J.~N.~Gruse, C.~I.~D.~Underwood, A.~F.~Antoine, C.~Arran, M.~Backhouse, C.~D.~Baird, M.~D.~Balcazar and N.~Bourgeois, \textit{et al.}
Nature Commun. \textbf{11}, no.1, 6355 (2020).


\bibitem{Aad:2020cws}
G.~Aad \textit{et al.} [ATLAS],
Phys. Rev. Lett. \textbf{125}, no.13, 131801 (2020).


\bibitem{Nicoli:2020njz}
K.~A.~Nicoli, C.~J.~Anders, L.~Funcke, T.~Hartung, K.~Jansen, P.~Kessel, S.~Nakajima and P.~Stornati,
Phys. Rev. Lett. \textbf{126}, no.3, 032001 (2021).

\bibitem{Tan21}
 D. -R. Tan, J. -H. Peng, Y. -H. Tseng, F. -J. Jiang, arXiv:2103.10846.  

\bibitem{kera}
https://keras.io

\bibitem{tens}
https://www.tensorflow.org



\bibitem{Wu82}
F.~Y.~Wu, Rev. Mod. Phys. {\bf 54}, 235 (1982).

\bibitem{Swe87}
R. H. Swendsen, and J.-S. Wang, (1987), Phys. Rev. Lett. {\bf 58(2)}, 86 (1987).


\bibitem{Wol89}
U. Wolff, 
Phys. Rev. Lett. {\bf 62}, 361 (1989).




\end{thebibliography}
\end{document}